\documentclass{ws-ijmpa}
\usepackage[super,compress]{cite}
\usepackage{graphicx}
\begin{document}

%
\catchline{}{}{}{}{}
%

\title{ON THE SCATTERING OF A HIGH-ENERGY COSMIC RAY ELECTRONS OFF THE DARK MATTER}

\author{V. Beylin
\footnote{344103 Sodruzhestva str.35/1 apt.149, Rostov-on-Don, Russia.}}

\address{Theoretical Physics Department, Southern Federal University, \\
344090 Stachki av. 194, Rostov-on-Don, Russia.\\
vitbeylin@gmail.com}

\author{M. Bezuglov}
	
\address{Moscow Institute of Physics and Technology (State University),\\
	9 Institutskiy per., 141701 Dolgoprudny, Moscow Region, Russia, \\
	Bogoliubov Laboratory of Theoretical Physics, Joint Institute for Nuclear Research,\\ 
	Joliot-Curie 6, 141980 Dubna, Moscow region, Russia \\
bezuglov.ma@phystech.edu}

\author{V. Kuksa}

\address{Theoretical Physics Department, Southern Federal University, \\
344090 Stachki av. 194, Rostov-on-Don, Russia\\
vkuksa47@mail.ru}

\author{E. Tretyakov}

\address{Department of Physics, Southern Federal University, \\
	344090 Zorge str.5, Rostov-on-Don, Russia\\
	Horoshome@gmail.com}

\author{A. Yagozinskaya}

\address{Department of Physics, Southern Federal University, \\
	344090 Zorge str.5, Rostov-on-Don, Russia\\
	Ali75622@gmail.com}

\maketitle

\begin{history}
\received{Day Month Year}
\revised{Day Month Year}
\end{history}

\begin{abstract}

High-energy cosmic ray electrons interaction with Dark Matter particles are considered. 
In particular, a weakening of energy spectrum of cosmic electrons is predicted resulting from inelastic electron scattering on hyper-pions in the hypercolor extension of the Standard Model. 
Corresponding cross section and angular distributions of secondary neutrino are calculated and studied. 
We also briefly discuss some effects of scattering processes of such type.

\end{abstract}

\ccode{PACS numbers: 14.80.Ec, 14.80.Bn, 12.60.Nz}

\section{Introduction}
The nature of the Dark Matter (DM) has been in the focus of fundamental physics attention for a long time. Attempts to penetrate this mysterious fortress from different directions are carried out persistently and regularly, using various tools, but a breach in the wall has not yet appeared.
The presence of objects contributing significantly to energy density of the Universe and manifesting themselves through gravitational interaction forces us to search for the Dark Matter signals emerging as a result of the DM annihilation or decay \cite{WIMPReview}. Of course, we are talking about indirect methods of the DM detection (see, for example, Refs.~\refcite{DMobserv1,DMobserv2,DMobserv3,DMobserv4,DMobserv5,DMastro,DMescattering}), unlike the direct finding of these (stable) particles in observations at the collider (LHC) or in scattering off nuclei in underground experiments\cite{LUX_1,Xenon_1,DARWIN_1}. 
Specifically, in a space there are diffuse or monochromatic fluxes of photons and/or leptons (in particular, neutrino) producing by the annihilating or decaying DM. However, the Universe is also permeated by cosmic ray streams consisting of protons, electrons, their antiparticles, light nuclei, photons originating from various sources such as processes in active Galaxy center, explosions of supernova and so on. Energies of these particles lie in a wide range - from keVs up to multi TeVs\cite{CR_spectrum_2008,DMobserv1}. It should be noted, photons of any energies can not move freely at the scale of the Galaxy and beyond because of intensive interaction with matter in contrast to neutrino.   

It seems reasonable to consider the processes of interaction of cosmic ray fluxes with the DM particles\cite{Profumo_Ubaldi_2011,CR_DM_indirect,CR_DM_indirect_2,DM_indirect_Review} forming the halo of the Galaxy\cite{DM_halo,DM_halo_2,DM_halo_3}. Indeed, such analysis can be useful to detect some peculiar signals that differ in energy spectrum or spatial distributions from the annihilation signals of the DM. More specifically, we calculate here  cross section of cosmic electron interaction with hyperpions (H-pions), which are the one of two DM component in vectorlike hypercolor extension of the Standard Model \cite{We_PR,PEPou,We_Adv}. As the model proposes, the DM consists of two pseudo-Nambu-Goldstone particles that are neutral and stable, but they interact with standard vector bosons and quarks via different ways \cite{We_Adv}. We will briefly discuss this point later.

Above mentioned process of high-energy electron inelastic scattering off the DM component, specifically, the neutral H-pion, results to production of two neutrinos from different vertexes. In the standard neutrino formation scheme, it is assumed that neutrino arise from meson decays. And the energy spectrum of the atmospheric neutrino is determined by the energies of cosmic rays and the type of meson that decays, creating a secondary neutrino. Namely, the decays of pions or muons occur through various channels, generating electronic or muon neutrinos with a steeper energy spectrum. It is also assumed that astrophysical neutrinos arise in collisions of nucleons and photonuclear reactions (with much smaller cross sections, which is partially compensated by a high photon density near astrophysical objects with high radiation activity and power, for example). Both these processes provide the bulk of high-energy astrophysical neutrinos (for more detail see Ref.~\refcite{GrefeNeutrinoDMdecay} and references therein). As we noted, the inelastic transition of electrons to high-energy neutrinos in interaction with the DM particles should also be important, despite the fact that the electrons make up only $\approx 1\, \%$ of cosmic rays in which protons dominate. The reason for our interest is that the secondary neutrino energy spectrum has an obvious feature - at high energies of incident electrons it practically copies the electron energy spectrum contrasting with the neutrino energy spectrum following from the meson decays.

In this paper we present some first results of a study of the inelastic interaction of cosmic rays with the DM particles in the framework of vectorlike hypercolor model. In the Section 2 we briefly describe basic elements of this model and of the analysis of the Dark Matter parameters. Then, the Section 3 is devoted to discussion of inelastic electron scattering off the H-pion DM component. In the Conclusion we summarize some results.

\section{Minimal vectorlike model and the DM carriers}

Here, we consider the minimal version of the SM extension by adding of the sector of additional fermions, hyperquarks (H-quarks), as it is used, for example, in Refs.~\refcite {San08,Kilic,SUNDRUM}. Initially, the simplest model with two H-quarks generations and two hypercolors, $N_ {HC} = 2$
was analyzed in Ref.~\refcite {We_PR} for the case of zero hypercharge. A comprehensive description of the procedure for construction of weak
interaction, starting from the standard-like chiral asymmetric
set of new fermion doublets with a nonzero hypercharge of H-quark generations is presented in Refs.~\refcite{PEPou,We_Adv}. As it has been shown there, two left doublets of
H-quarks can be transformed into one doublet of Dirac H-quarks with vectorlike weak interaction to avoid troubles of ``standard'' technicolor.
 Importantly, hypercharges of H-quark generations should have the same values and opposite signs to enforce the absence of
anomalies in the model. Notice at once, that the H-quark masses are degenerate, $M_U = M_D$, at the one loop level as it follows from the cancellation of the self-energy contributions of electroweak and H-pion loops which are exactly the same for both quarks. 

To form the Dirac states which correspond to
constituent quarks, it is used a scalar field with non-zero
vacuum expectation value (v.e.v.). This field (hyper-$\sigma-$ meson) is introduced as
a scalar singlet pseudo-Nambu--Goldstone (pNG) boson in the framework of the simplest
linear sigma-model. The structure of the
pNG multiplet in this minimal extension is defined by the global
symmetry breaking $SU(4)\to Sp(4)$. The
Lagrangian has a specific global $U_{HB}(1)$ symmetry providing stability of the lightest neutral H-baryon/H-diquark states ($B^0, \,\,\bar B^0$) possessing an additive conserving H-baryon number. At the same time, the lightest neutral H-pion state is stable due to conserving of multiplicative modified charge conjugation (hyper- G or HG)-parity\cite{We_Adv}. 

Complete set of the lightest spin-0 H-hadrons in the model
includes pNG states (pseudoscalar H-pions $\tilde\pi_k$ and scalar
complex H-diquarks/H-baryons $B^0$), their opposite-parity chiral partners
$\tilde a_k$ and $A^0$, and singlet H-mesons $\tilde\sigma$ and
$\tilde\eta$. These states correspond to H-quark currents with different quantum numbers, all of them are listed in Table \ref{tab:H-hadrons} where $\tilde G$ denotes hyper-$G$-parity of a state, $\tilde B$ is the H-baryon number. $Q_\text{em}$ is the electric charge, $T$ is the weak isospin. Notice, H-baryons have not intrinsic $C$- and $HG$-parities, because of the charge conjugation reverses the sign of the H-baryon number.. 
The model suggested contains the elementary Higgs field which is not a pNG state. This is enough to consider those processes of inelastic electron scattering which we are interested on. 
\begin{table}[ph]
	\tbl{Quantum numbers of the lightest (pseudo)scalar H-hadrons and H-quark currents in $SU(2)_{HC}$ model.}
	{\begin{tabular}{@{}ccccccccc@{}}
		\hline
		state & $$ & H-quark current & $$ & $T^{\tilde G}(J^{PC})$ & $$ & $\tilde B$ & $$ & $Q_\text{em}$ \\
		\hline
		$\tilde\sigma$ & $$ & $\bar Q Q$ & $$ & $0^+(0^{++})$ & $$ & 0 & $$ & 0 \\
		$\tilde\eta$ & $$ & $i \bar Q \gamma_5 Q$ & $$ & $0^+(0^{-+})$ & $$ & 0 & $$ & 0 \\
		$\tilde a_k$ & $$ & $\bar Q \tau_k Q$ & $$ & $1^-(0^{++})$ & $$ & 0 & $$ & $\pm 1$, 0 \\
		$\tilde\pi_k$ & $$ & $i \bar Q \gamma_5 \tau_k Q$ & $$ & $1^-(0^{-+})$ & $$ & 0 & $$ & $\pm 1$, 0 \\
		$A^0$ & $$ & $\bar Q_{a\underline a}{}^{C} \epsilon_{ab} \epsilon_{\underline a \underline b} Q_{b\underline b}$ & $$ & $0^{\hphantom{+}}(0^{-\hphantom{+}})$ & $$ & 1 & $$ & 0 \\
		$B^0$ & $$ & $i \bar Q_{a\underline a}{}^{C} \epsilon_{ab} \epsilon_{\underline a \underline b} \gamma_5 Q_{b\underline b}$ & $$ & $0^{\hphantom{+}}(0^{+\hphantom{+}})$ & $$ & 1 & $$ & 0 \\
		\hline
			\end{tabular} \label{tab:H-hadrons}}
\end{table}

We consider above mentioned neutral pNG particles as the the DM carriers analogously to Refs.~\refcite{San08,DMou1,DMou2,DMou3}. To discuss more definitely some processes with them, we represent here that parts of physical Lagrangian which are relevant for analysis of stable H-pion scenario\cite{Bai,We_Adv}. 

The H-quark interactions
with the EW bosons are vectorlike, and the corresponding Lagrangian has the following form:
\begin{align}\label{3.1}
L(Q,G)={}&\frac{1}{\sqrt{2}}g_W\bar{U}\gamma^{\mu}D
W^+_{\mu}+\frac{1}{\sqrt{2}}g_W\bar{D}\gamma^{\mu}U W^-_{\mu}\notag\\
&+\frac{1}{2}g_W(\bar{U}\gamma^{\mu}U-\bar{D}\gamma^{\mu}D)(c_W
Z_{\mu}+s_W A_{\mu}).
\end{align}
Here $U,\,\, D$ are H-quark fields, $c_W$ and $s_W$ denote cosine and sine of the Weinberg angle.
Interactions of (pseudo)scalars with photons and intermediate bosons
are described by Lagrangians:
\begin{align}\label{eq:SG1}
{L}(\tilde\sigma,H,G)
= \frac18 \left[ 2 g_W^2 W_\mu^+ W^\mu_- + ( g_B^2 + g_W^2 ) Z_\mu Z^\mu \right] (\cos\theta_s H - \sin\theta_s \tilde \sigma)^2,
\end{align}

\begin{align}\label{eq:SG2}
L(\tilde\pi,\tilde a, G) = & \left[  i g_W W_+^\mu \left( \tilde\pi^0  \tilde\pi^-_{,\mu}- \tilde\pi^-
\tilde\pi^0_{,\mu} \right) + \text{h.c.} \right] + i g_W ( c_W
Z^\mu - s_W A^\mu ) (\tilde\pi^- \tilde\pi^+_{,\mu}-\tilde\pi^+
\tilde\pi^-_{,\mu})
\notag\\
&+ g_W^2 \tilde \pi^+ \tilde\pi^- ( c_W Z^\mu - s_W A^\mu )^2
- g_W^2 \tilde\pi^0 ( c_W Z^\mu - s_W A^\mu ) \left( \tilde\pi^+ W^-_\mu + \tilde\pi^- W^+_\mu \right)
\notag\\
&-\frac12 g_W^2 \left( \tilde\pi_+^2 W^-_\mu W_-^\mu  + \tilde\pi_-^2 W^+_\mu W_+^\mu \right) + g_W^2
\left( \tilde\pi_0^2+\tilde\pi^- \tilde\pi^+ \right) W^+_\mu W_-^\mu +(\tilde\pi \to \tilde a)  .
\end{align}
In the Lagrangian $L(\tilde\pi,\tilde a, G)$ the last term means that the interactions of the triplet of scalar H-mesons $\tilde a$ have the same couplings and vertexes as the interactions of H-pions. 

The fields $\tilde\sigma$, $\tilde\pi$, $H$ (here $H$ is the Higgs boson field) interact with the H-quarks as it
is described by the following Lagrangian:
\begin{align}\label{3.3}
L(Q,\tilde\sigma,H)={}&-\kappa (c_{\theta} \tilde{\sigma}+s_{\theta}
H)(\bar{U}U+\bar{D}D)+i\sqrt{2}\kappa \tilde{\pi}^+\bar{U}\gamma_5
D\notag\\&+i\sqrt{2}\kappa \tilde{\pi}^-\bar{D}\gamma_5 U+i\kappa
\tilde{\pi}^0(\bar{U}\gamma_5 U-\bar{D}\gamma_5 D),
\end{align}
where $c_\theta= \cos\theta$ and $s_\theta= \sin\theta$.

The Lagrangian for self-interactions of scalar fields can be found in Ref.~\refcite{We_Adv}  and it can be useful for demonstration of some specific channels of interactions of two DM components. Consideration of these processes is beyond the scope of the paper, so we will omit this part of Lagrangian here.

It is important, all restrictions on the oblique corrections are
fulfilled in this variant of hypercolor \cite{We_PR, PEPou}. In the scenario with a non-zero hypercharge and $h$--$\tilde{\sigma}$ mixing a constraint for the $T$ parameter
value emerges (see Refs.~\refcite{We_PR,DMou1}). Here ${\theta}$ is the angle of mixing between H-sigma and the Higgs boson which controls the consistency of the model predictions with the Standard Model precision measurements. Its value is estimated from analysis of Peskin-Tackeuchi (PT) parameters. It has been found that  $S_{\theta}\equiv \sin\theta \lesssim 0.1$ (see Ref.~\refcite{We_PR}) to
get away problems with the PT parameters and the measured characteristics of the SM Higgs boson.
Then, to analyze quantitavely processes involving the DM particles, it is necessary to know only a few parameters. Specifically, these are tree-level masses of the DM components (H-pion and H-baryon) and mass and v.e.v. of $\tilde \sigma-$ meson.

First of all, it was necessary to confirm that the neutral component of the H-pion triplet is the lightest. The mass difference in this triplet results from electroweak contributions only and is well known: $\Delta M_{\tilde \pi}= m_{\tilde{\pi}^{\pm}}-m_{\tilde{\pi}^0} \approx 0.16 \,\,\mbox{GeV}$, so charged H-pion states can decay producing neutral H-pion (more detail can be found in Ref.~\refcite{We_Adv}). Importantly that  non-zero mass splitting in the
H-pion triplet violates isotopic invariance, however, $HG$-parity
remains a conserved quantum number since it corresponds to a discrete
symmetry. Thus, the neutral H-pion remains stable independently on higher order corrections.
 
And the mass splitting $\Delta M_{B-\tilde \pi} = m_{B^0}-m_{\tilde{\pi}^0}$ is determined, as in the triplet of H-pions, only by electroweak diagrams due to mutual cancellation of all other contributions. However, the somewhat different origin of these neutral and stable particles results to the following expression depending on the renormalization scale:
\begin{align} 
\Delta M_{B-\tilde \pi}=\frac{-g^2_2m_{\tilde{\pi}}}{16\pi^2}\Bigg[ 8\beta^2-1-(4\beta^2-1)\ln\frac{m_{\tilde{\pi}}^2}{\mu^2}+2\frac{M_W^2}{m_{\tilde{\pi}}^2}\left(\ln\frac{M_{W}^2}{\mu^2}-\beta^2 \ln\frac{M_W^2}{m_{\tilde{\pi}}^2}  \right)
\nonumber \\ 
-8\frac{M_W}{m_{\tilde{\pi}}}\beta^3\left(\arctan\frac{M_W}{2m_{\tilde{\pi}}\beta}+\arctan\frac{2m_{\tilde{\pi}}^2-M_W^2}{2m_{\tilde{\pi}}M_W\beta} \right)
\bigg],
\end{align}
where $\beta=\displaystyle \sqrt{1-\frac{M_W^2}{4m_{\tilde{\pi}}^2}}$. Dependence of the mass splitting on the renormalization point is a consequence of the coupling of these states with different H-quark currents that is these (possible) components of the DM are produced by different mechanisms. In any case, we need to consider dependence of the DM measurable parameters on the renormalization parameter value. 

We also suppose that other (not pNG) possible H-hadrons including vector H-mesons are heavier than the pNG bosons. In other words, the scale of the explicit $SU(4)$ symmetry breaking  is small in comparison with the scale of the dynamical symmetry breaking. 
It is an analogy with the QCD, where the scale of chiral symmetry breaking is much larger than the masses of light quarks. So, we assume that the masses
of low-lying H-states are of the order of $10^3\, \mbox{GeV}$. The next step to evaluate  possible values of these masses is the studying of them as the DM carriers.  

Remind, this minimal hypercolor scenario has the specific symmetry resulting from the invariance of H-quark fields under hyper-$G$-parity (see Refs.~\refcite{Bai,We_Adv} and references therein). As a consequence, there arise the following channels of  $\tilde{\pi}^{\pm}$ decay: $\tilde{\pi}^{\pm}\to \tilde{\pi}^0 \pi^{\pm}$ and $\tilde{\pi}^{\pm}\to\tilde{\pi}^0 l^{\pm}\nu_l$. Expressions for the decay widths can be found in Ref.~\refcite{We_Adv} and numerically we get
\begin{align} \label{Br}
\Gamma(\tilde{\pi}^{\pm}\to\tilde{\pi}^0 l^{\pm}\nu_l)&=6\cdot
10^{-17}\,\mbox{GeV},\,\,\,\tau_{l}=1.1\cdot10^{-8}\,\mbox{sec};\notag\\
\Gamma(\tilde{\pi}^{\pm}\to\tilde{\pi}^0 \pi^{\pm} )&=3\cdot10^{-15}\,\mbox{GeV},\,\,\,\tau_{\pi}=2.2\cdot10^{-10}\,\mbox{sec}.
\end{align}
Now we can consider some features of the two-component Dark Matter in more detail.  

\section{Two-component Dark Matter in the vectorlike H-color model}
In fact, there are five Boltzmann kinetic equations, since we must take into account the two states of the neutral H-baryon, $B^0, \, \, \bar B ^ 0$ and two charged H-pions together with the neutral one. The reason is that the mass splitting in the triplet of H-pions is very small, so the processes of co-annihilation\cite{GriestSeckel} contribute significantly to the annihilation cross section. Numerically it has been shown in another vectorlike scenario in Ref.~\refcite {DMou3}. 

So, we start from the system of five Boltzmann equations(\ref{Boltzman1a})-(\ref{Boltzman1b}) \footnote{We neglect here by forward and backward reactions of type $iX\rightarrow jX$ which are not important for this analysis.}, one for each DM component($i,j = \tilde{\pi}^+,\,\tilde{\pi}^-,\tilde{\pi}^0;\mu, \nu =B,\, \bar{B}$):    
\begin{eqnarray} 
\label{Boltzman1a}
\frac{da^3n_i}{a^3dt}=-\sum\limits_j<\sigma v>_{ij}\left(n_in_j-n_i^{eq}n_j^{eq}\right)-\sum\limits_j\Gamma_{ij}\left(n_i-n_i^{eq}\right)-
\nonumber \\
-\sum\limits_{j,\mu,\nu}<\sigma v>_{ij\rightarrow \mu\nu}\left(n_in_j-\frac{n_i^{eq}n_j^{eq}}{n_{\mu}^{eq}n_{\nu}^{eq}}n_{\mu}n_{\nu}\right)+ \notag \\
\sum\limits_{j,\mu,\nu}<\sigma v>_{\mu\nu\rightarrow ij}\left(n_{\mu}n_{\nu}-\frac{n_{\mu}^{eq}n_{\nu}^{eq}}{n_{i}^{eq}n_{j}^{eq}}n_{i}n_{j}\right),
\end{eqnarray}
\begin{eqnarray}
\label{Boltzman1b}
\frac{da^3n_{\mu}}{a^3dt}=-\sum\limits_{\nu}<\sigma v>_{\mu\nu}\left(n_{\mu}n_{\nu}-n_{\mu}^{eq}n_{\nu}^{eq}\right)+ \notag \\
\sum\limits_{\nu,i,j}<\sigma v>_{ij\rightarrow \mu\nu}\left(n_in_j-\frac{n_i^{eq}n_j^{eq}}{n_{\mu}^{eq}n_{\nu}^{eq}}n_{\mu}n_{\nu}\right)-\nonumber \\
\sum\limits_{\nu,i,j}<\sigma v>_{\mu\nu\rightarrow ij}\left(n_{\mu}n_{\nu}-\frac{n_{\mu}^{eq}n_{\nu}^{eq}}{n_{i}^{eq}n_{j}^{eq}}n_{i}n_{j}\right),
\end{eqnarray}
where: 
\begin{align} 
<\sigma v>_{ij}=<\sigma v>(ij\rightarrow XX) \notag \\
<\sigma v>_{ij\rightarrow \mu\nu}=<\sigma v>(ij\rightarrow \mu\nu) \notag \\
\Gamma_{ij}=\Gamma(i\rightarrow jXX),
\end{align}
and analogously for $\mu$ and $\nu$ components.

All charged H-pions will eventually decay into $\tilde \pi^0$ as it was noted above, so the main parameter is the total density of $\tilde \pi $ particles, $n_{\tilde \pi}=\sum_{i}n_i$. The $B^0$ and $\bar{B^0}$ particles are stable and we also introduce and consider their  total density $n_{B}=\sum_{\mu}n_{\mu}$. 
Using the notations above and the approximation $n_i/n=n_i^{eq}/n^{eq}$ which can be used  for co-annihilation, we rewrite the system of equations in the following form, (\ref{Boltzman2a})-(\ref{Boltzman2b}): 
\begin{eqnarray} 
\label{Boltzman2a}
\frac{da^3n_{\pi}}{a^3dt}=\bar{<\sigma v>}_{\tilde \pi}\left(n_{\tilde \pi}^2-\left(n_{\tilde \pi}^{eq}\right)^2\right)
-<\sigma v>_{\tilde \pi \tilde \pi}\left(n_{\tilde \pi}^2-\frac{\left(n_{\tilde \pi}^{eq}\right)^2}{\left(n_{B}^{eq}\right)^2}n_{B}^2\right)+ \nonumber \\
<\sigma v>_{BB}\left(n_{B}^2-\frac{\left(n_{B}^{eq}\right)^2}{\left(n_{\tilde \pi}^{eq}\right)^2}n_{\tilde \pi}^2\right),
\\
\label{Boltzman2b}
\frac{da^3n_{B}}{a^3dt}=\bar{<\sigma v>}_{B}\left(n_{B}^2-\left(n_{B}^{eq}\right)^2\right)
+<\sigma v>_{\tilde \pi \tilde \pi}\left(n_{\tilde \pi}^2-\frac{\left(n_{\tilde \pi}^{eq}\right)^2}{\left(n_{B}^{eq}\right)^2}n_{B}^2\right)- \nonumber \\
<\sigma v>_{BB}\left(n_{B}^2-\frac{\left(n_{B}^{eq}\right)^2}{\left(n_{\tilde \pi}^{eq}\right)^2}n_{\tilde \pi}^2\right),
\end{eqnarray}
where:
\begin{eqnarray} 
\bar{<\sigma v>}_{\tilde \pi}=\frac{1}{9}\sum\limits_{i,j}<\sigma v>_{ij},~~~~\bar{<\sigma v>}_{B}=\frac{1}{4}\sum\limits_{\mu,\nu}<\sigma v>_{\mu\nu},
\notag \\
<\sigma v>_{\tilde \pi \tilde \pi}=\frac{1}{9}(<\sigma v>(\tilde \pi^0 \tilde \pi^0\rightarrow B\bar{B})+2<\sigma v>(\tilde \pi^+\tilde \pi^-\rightarrow B\bar{B})),
\notag \\
<\sigma v>_{BB}=\frac{1}{2}(<\sigma v>(B\bar{B}\rightarrow\tilde \pi^0 \tilde \pi^0)+<\sigma v>(B\bar{B}\rightarrow\tilde \pi^-\tilde \pi^+)).
\end{eqnarray}
Further, we can simplify the system (\ref{Boltzman2a})-(\ref{Boltzman2b}) assuming that $m_{\tilde \pi }/M_B \approx 1$. Then $n_{B}^{eq}/n_{\tilde \pi}^{eq}=2/3$ and we have:
\begin{eqnarray} 
\label{Boltzman3a}
\frac{da^3n_{\tilde \pi}}{a^3dt}=\bar{<\sigma v>}_{\tilde \pi}\left(n_{\tilde \pi}^2-\left(n_{\tilde \pi}^{eq}\right)^2\right)
-<\sigma v>_{\tilde \pi \tilde \pi}\left(n_{\tilde \pi}^2-\frac{9}{4}n_{B}^2\right)+ \nonumber \\
<\sigma v>_{BB}\left(n_{B}^2-\frac{4}{9}n_{\tilde \pi}^2\right),
\\
\label{Boltzman3b}
\frac{da^3n_{B}}{a^3dt}=\bar{<\sigma v>}_{B}\left(n_{B}^2-\left(n_{B}^{eq}\right)^2\right)
+<\sigma v>_{\tilde \pi \tilde \pi}\left(n_{\tilde \pi}^2-\frac{9}{4}n_{B}^2\right)- \nonumber \\
<\sigma v>_{BB}\left(n_{B}^2-\frac{4}{9}n_{\tilde \pi}^2\right).
\end{eqnarray}
Here, we consider the case when the mass splitting between $m_{\tilde \pi^0}$ and $M_{B^0}$ is not large $\Delta M_{B^0-\tilde \pi^0}|/m_{\tilde \pi^0} \lesssim 0.02$. Thus, the cross sections $<\sigma v>_{\tilde \pi \tilde \pi}$ and $<\sigma v>_{BB}$ should be calculated taking into account the temperature dependence as it should be for any  process which occurs near the threshold\cite{GriestSeckel}: 
\begin{eqnarray} 
<\sigma v>_{BB}\approx <(a+bv^2)v_2>=\frac{2}{\sqrt{\pi}x}\left(a+\frac{8b}{x}\right),
\end{eqnarray}
where $x=m_{\tilde \pi}/T$ and $v_2$ is the velocity of final particles in the center-of-mass frame. 

In order to solve the system (\ref{Boltzman3a})-(\ref{Boltzman3b}) we use standard notations: $Y=n/s$ and $x=m_{\tilde \pi}/T$, where $s$ is the entropy density. So, we get\footnote {We neglect terms $~\Delta M_{\tilde \pi}/M_{B}$ due to the small mass splitting between neutral components of the DM.}:
\begin{eqnarray} 
\label{Boltzman4a}
\frac{dY_{\pi}}{dx}=g(x,T)\cdot \left[\lambda_{\tilde \pi}((Y_{\tilde \pi}^{eq})^2-Y^2_{\tilde \pi})-\lambda_{\tilde \pi \tilde \pi}\left(Y_{\tilde \pi}^2-\frac{9}{4}Y_B^2\right)+ \lambda_{BB}\left(Y_{B}^2-\frac{4}{9}Y_{\tilde \pi}^2\right)\right]
\\
\label{Boltzman4b}
\frac{dY_{B}}{dx}= g(x,T) \cdot \left[\lambda_{B}((Y_{B}^{eq})^2-Y^2_{B})+\lambda_{\tilde \pi \tilde \pi}\left(Y_{\tilde \pi}^2-\frac{9}{4}Y_B^2\right)- \lambda_{BB}\left(Y_{B}^2-\frac{4}{9}Y_{\tilde \pi}^2\right)\right]
\end{eqnarray}
and $g(x,T)=\displaystyle \frac{\sqrt{g(T)}}{x^2}\left\{1+ \displaystyle \frac{1}{3}\frac{d(\log g(T))}{d(\log T)}\right\}$.
Here, we use notations from Ref.~\refcite{Steigman}:  $\lambda_i=2.76\times 10^{35}m_{\tilde \pi}<\sigma v>_{i}$ ($m$ is in GeV and $<\sigma v>$ in $cm^3s^{-1}$) and $Y_{\tilde \pi}^{eq}=0.145(3/g(T))x^{3/2}e^{-x},~ Y_{B}^{eq}=0.145(2/g(T))x^{3/2}e^{-x}$, $g(T)$ is the number of relativistic degrees of freedom contributing to the energy density\footnote{Generally speaking, we should to distinguish between relativistic degrees of freedom contributing to the energy density($g_{\rho}$) and the ones who determine the total entropy density($g_{s}$), however,  in the Universe this distinguishing occurs only after annihilation into photons of all electron-positron pairs and it happens a long time after the DM relic formation.}.
Note that the function $g(T)$ can be effectively approximated as:
\begin{eqnarray} 
g(T)\simeq \frac{115}{2}+\frac{75}{2}\tanh\left[2.2\left(\log_{10}T+0.5\right)\right]+10\tanh\left[3\left(\log_{10}T-1.65\right)\right],
\end{eqnarray}
and for numerical analysis we use this formula instead of known estimation $g(T)\approx 100$. 

The present relic density $\Omega h^2$ can be written in terms of the relic abundance $\rho$ and critical mass density $\rho_{crit}$
\begin{eqnarray} 
\Omega h^2 = \frac{\rho}{\rho_{crit}}h^2=\frac{m s_0 Y_0 }{\rho_{crit}}h^2\simeq 0.3 \times 10^9 \frac{m}{GeV} Y_0.
\end{eqnarray}
The subscript "0" denotes quantities whose values are evaluated at present time.

To solve the system (\ref{Boltzman4a})-(\ref{Boltzman4b}) numerically, it is convenient to make the replacement \cite{Steigman} $W=\log Y$.
 Some of these solutions will be presented below as the set of regions in the plane of parameters, i.e. H-pion and H-sigma masses. (Remind, the  relation between these masses depends also on the mixing angle $\theta$ and renormalization scale $\mu$; moreover, the dependence on the vacuum parameter $u$ is also taken into account.)

For better understanding, we indicate physically interesting areas by different hatching. Namely, the hatching with vertical cells denotes areas where we have correct DM relic density\footnote{This area is extended to the values of accuracy of three sigma.}, in these regions fraction of H-pions is less than 25 percents ($0.1047 \le \Omega h^2_{HP}+\Omega h^2_{HB} \le 0.1228$ and $\Omega h^2_{HP}/(\Omega h^2_{HP}+\Omega h^2_{HB}) \le 0.25$). The hatching with oblique cells indicates domains where all parameters are exactly the same, but here H-pions make up just over a quarter of the DM ($0.1047 \le \Omega h^2_{HP}+\Omega h^2_{HB} \le 0.1228$ and $0.25\le \Omega h^2_{HP}/(\Omega h^2_{HP}+\Omega h^2_{HB}) \le 0.4$). Importantly, we do not have any areas where H-pion component can dominate in the Dark Matter. The reason is obvious: H-pions have much more channels of interaction via weak vector bosons, i.e. chances of  their ``burnout'' more than the other, $B^0$, component has. In contrast to neutral H-pions, $B^0$ mesons interact with the world of ordinary particles only via H-quark and H-pion loops at the first nonzero order.

Further,  hatching with horizontal lines denotes areas which correspond to permitted regions ($ \Omega h^2_{HP}+\Omega h^2_{HB} \le 0.1047$) and here we can not explain the DM relic abundance only by H-color components. Regions that are hatched with vertical lines are forbidden by direct experiments of the XENON collaboration (see also Refs.~\refcite{Xenon_1,DARWIN_1}).

\begin{figure}[h]{}
	\centering
	\includegraphics[width=6cm,clip]{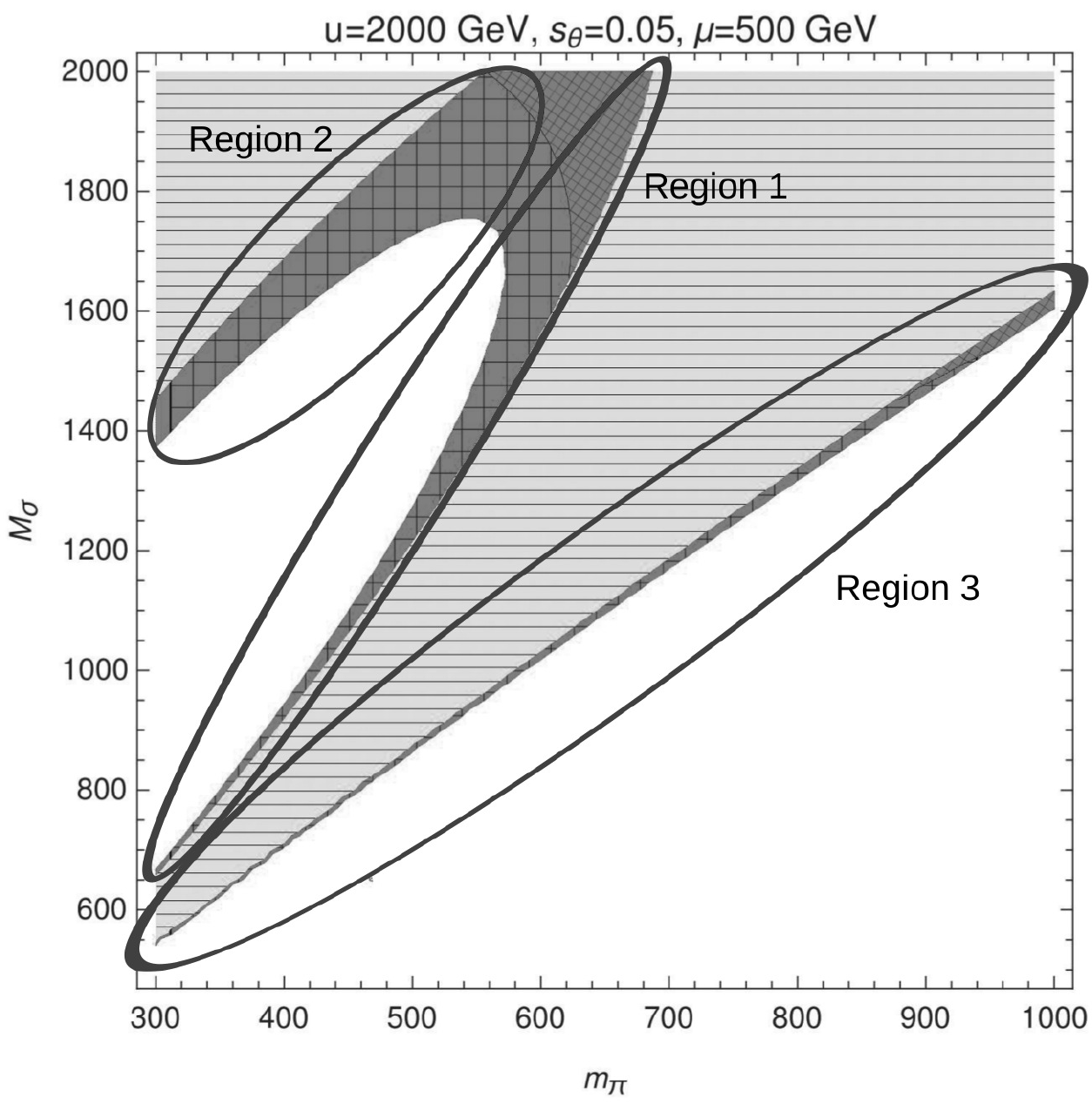}
	\caption{Numerical solution of the kinetic equations  system in a phase diagram in terms of $M_{\tilde \sigma}$ and $m_{\tilde \pi}$ parameters;  types of hatching are indicated in the text above.}
	\label{Regions123}
\end{figure}

So,  as it can be seen in Fig.~\ref{Regions123}, there are three areas where the recent DM density can be explained by the H-color model, in particular:

{\bf Region 1}: $M_{\tilde{\sigma}}>2m_{\tilde \pi^0}$ and $u \ge M_{\tilde{\sigma}} $. At small angles of mixing, $S_{\theta}$, and large masses of H-pions it is possible to obtain a good fraction of H-pions. 

{\bf Region 2}: the same relation between $M_{\tilde{\sigma}}, \,\, m_{\tilde \pi^0},\,\, u$ but the H-pion mass is smaller, $m_{\tilde \pi}\approx 300-600 \,\, \mbox{GeV}$ . Here the H-pion fraction is small.

{\bf Region 3}: $M_{\tilde{\sigma}}<2m_{\tilde \pi}$. This region is always possible and it can be visible in all figures. Note, here the process $\tilde \sigma \to \tilde \pi \tilde \pi$ is obviously absent, so, the two-photon signal from reaction $pp\to \tilde \sigma \to \gamma\gamma X$ could be, in principle, detected at the LHC. The H-pion fraction in the DM relic can be large if the mass $m_{\tilde \pi^0}$ is large and the mixing angle is small. 
   
\begin{figure}[h]
	\begin{minipage}[h]{0.43\linewidth}
		\center{\includegraphics[width=0.7\linewidth]{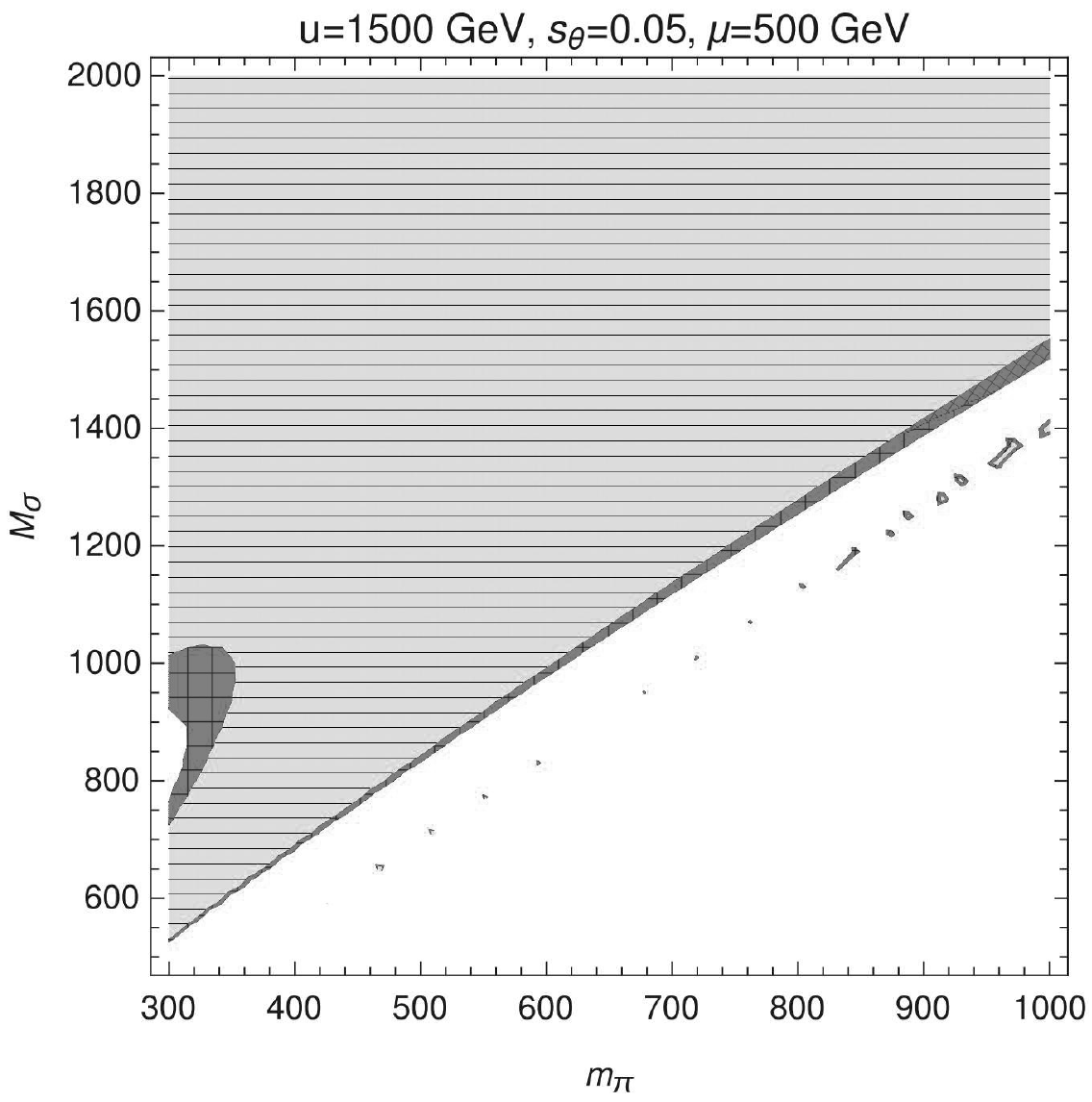}}
		\caption{Perhaps, a small "islands" should be a line of valid values of parameters.}
	\end{minipage}
	\hfill
	\begin{minipage}[h]{0.43\linewidth}
		\center{\includegraphics[width=0.7\linewidth]{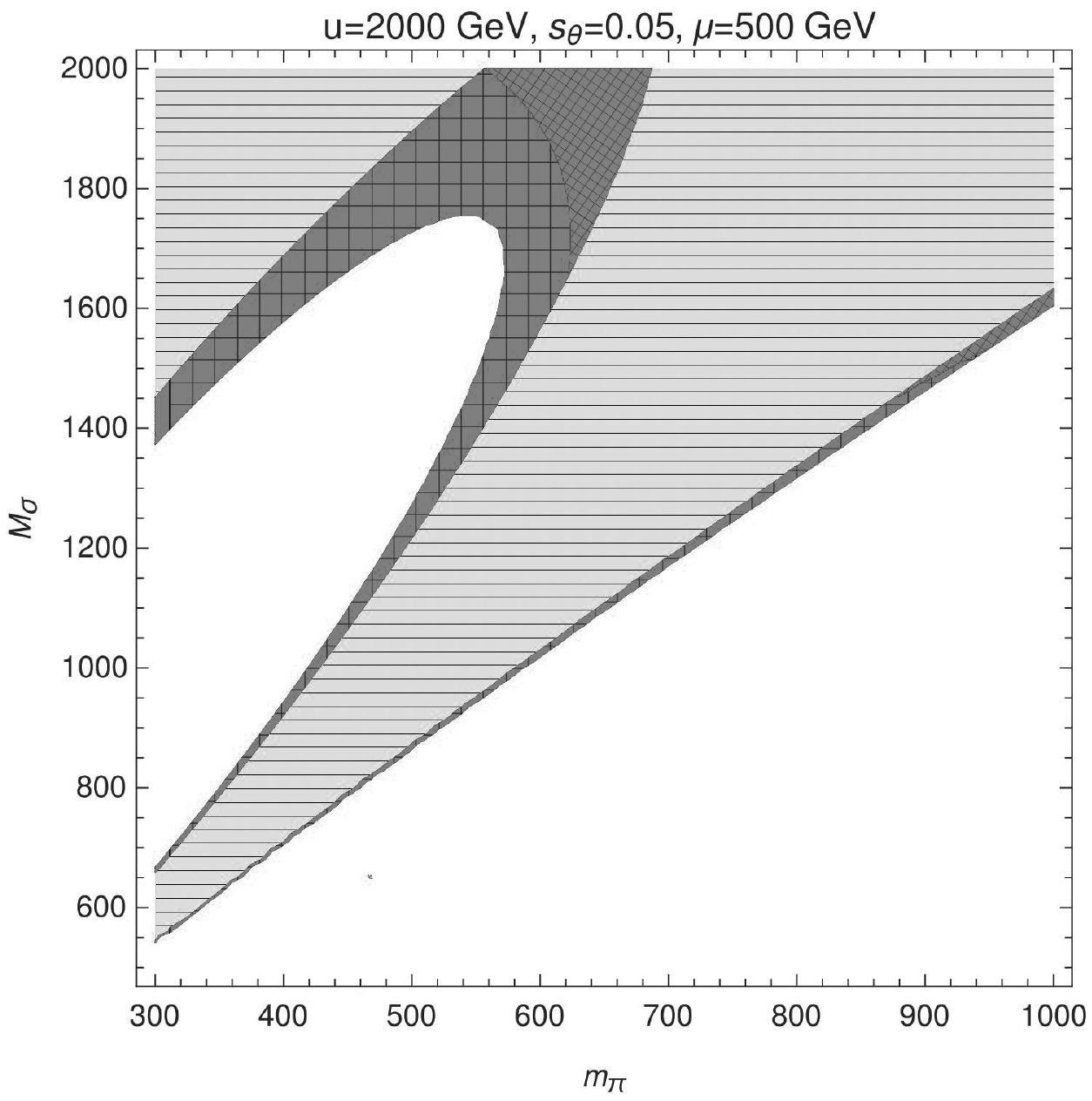}}
		\caption{The same as in Fig.1}
	\end{minipage}
	\vfill
	\begin{minipage}[h]{0.43\linewidth}
		\center{\includegraphics[width=0.7\linewidth]{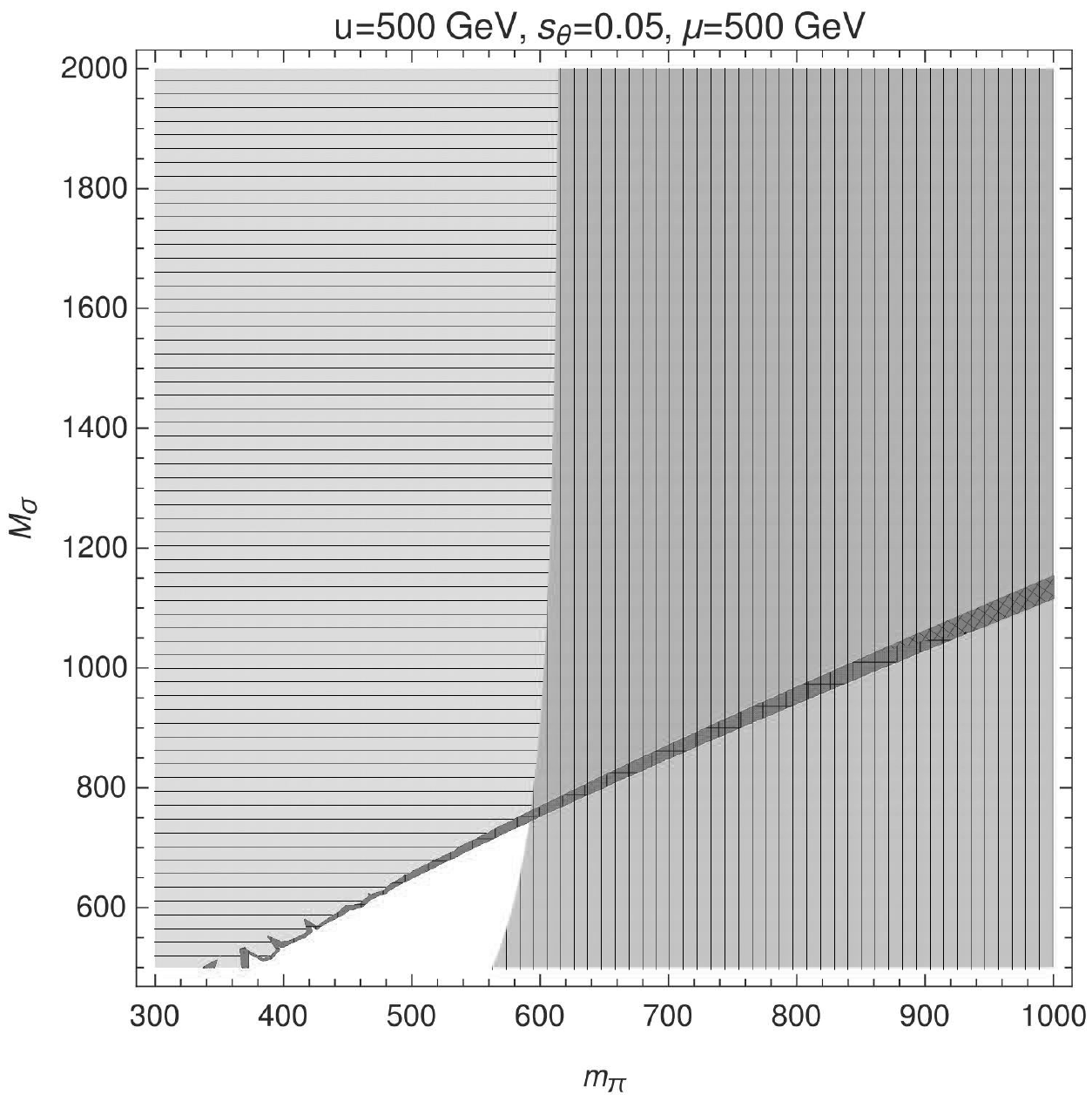}}
		\caption{Some forbidden areas are shown here.}
	\end{minipage}
	\hfill
	\begin{minipage}[h]{0.43\linewidth}
		\center{\includegraphics[width=0.7\linewidth]{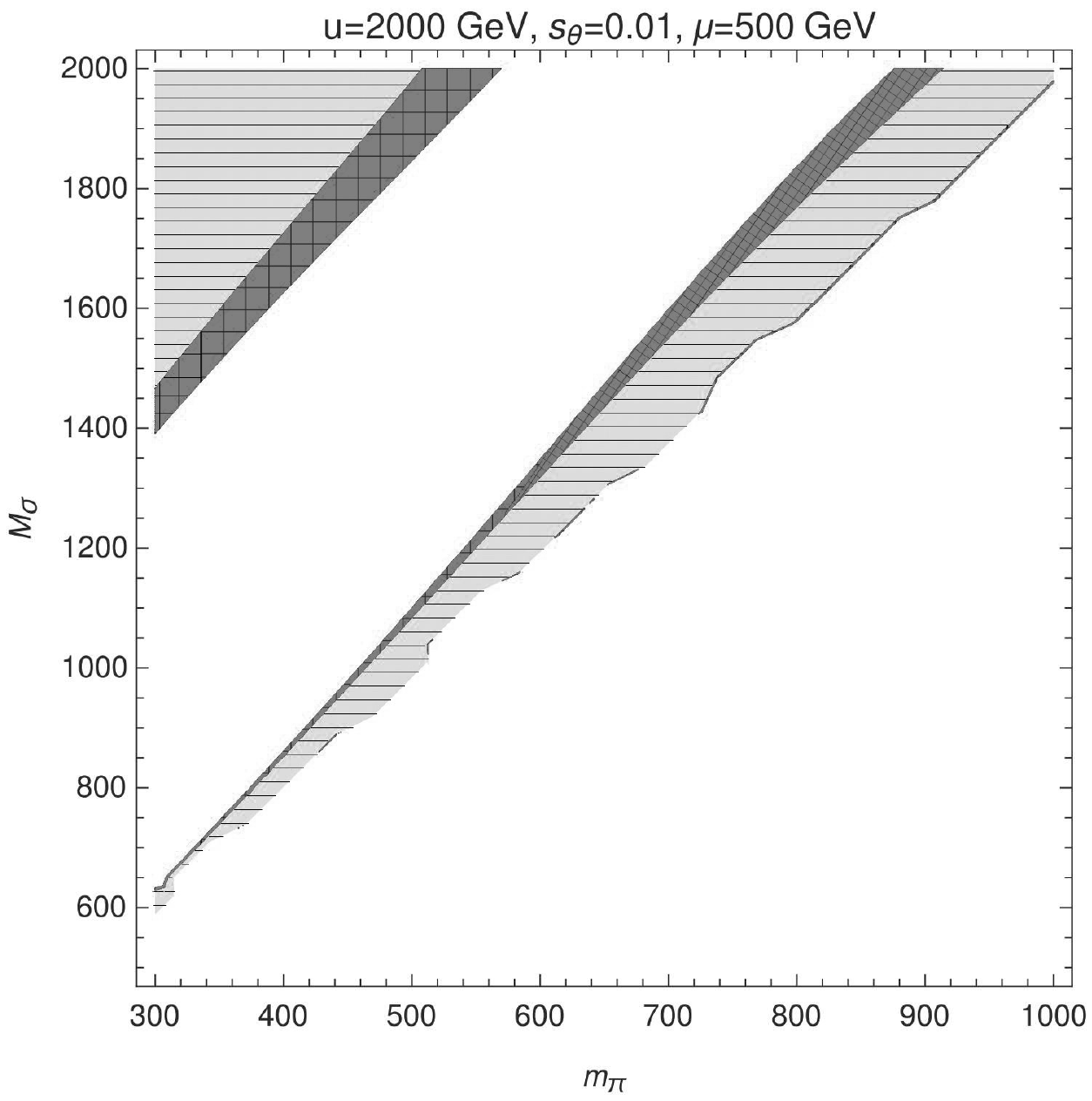}}
		\caption{A slightly changed numerical parameters in comparison with Fig. 2.}
	\end{minipage}
	\vfill
\end{figure}
A set of phase diagrams in terms of $M_{\tilde \sigma}$ and $m_{\tilde \pi}$ for some other numerical solutions is shown in Figs. 2-5 with the same designations. It can be concluded now that there are some values of the DM carriers masses which can be used in forthcoming analysis of the DM interactions with cosmic ray particles. 

\section{The cosmic electron scattering off H-pions}

Now, having in hands a reasonable estimations of the DM particles masses which are resulted from solution of the basic kinetic equations, we consider the scattering of high energy cosmic electrons on the DM\cite{DM_e_2,DM_e_3}. This problem is tightly connected with the studying both of peculiarities of cosmic neutrino fluxes\cite{DM_nu_1,DM_nu_2,DM_nu_3} and spectra of cosmic electrons (positrons)\cite{DM_e_1}, as they are determined by processes of the DM annihilation or decay\cite{GrefeNeutrinoDMdecay,DM_ann_decay}. 
We are interested here in analysis of high-energy neutrino production in the reaction $e \tilde \pi^0 \to \nu_e \tilde \pi^-$, the final $\tilde \pi^-$ state  
must decay according to the channels described above. Here, to estimate the total cross section we will use simple approximation: 
$\sigma(e \tilde \pi^0 \to \nu_e \tilde \pi^0 l \nu'_l ) \approx \sigma((e \tilde \pi^0 \to \nu_e \tilde \pi^-)\cdot Br( \tilde \pi^-\to \tilde \pi^0 l \nu'_l )$. Moreover, as it follows from the expressions (\ref{Br}) $Br( \tilde \pi^-\to \tilde \pi^0 e \nu'_e) \approx 0.01$ and $Br( \tilde \pi^-\to \tilde \pi^0 \pi^-) \approx 0.99$. Because we consider here  final $\tilde \pi^-$ as close to its mass shell, the charged standard pion decays into $e \nu_e$ and $\mu \nu_{\mu}$ with substantially different probabilities, namely $\approx 1.2\cdot 10^{-6}$ and $\approx 0.999$, correspondingly. From the decay $\tilde \pi^-\to \tilde \pi^0 l \nu'_l$  in the channel with intermediate $W-$ boson we get final lepton states 
$e \nu_e$ and $\mu \nu_{\mu}$ with equal probabilities. So, the scheme of this reaction is such: energetic cosmic electron produce the secondary electronic neutrino in the vertex $We \nu_e$ and secondary particles $e' \nu'_e$ or $\mu \nu_{\mu}$ arise from different decay channels of $\tilde \pi^-$. With an accounting of all branchings, we come to final states with $Br(1)=Br(\tilde \pi^0 \nu_e \mu'\nu'_{\mu})\approx 0.99$ and $Br(2)=Br(\tilde \pi^0 \nu_e e'\nu'_e)\approx 10^{-2}$. The statements above can be deduced more accurately from factorization approach for amplitudes of the such type\cite{Kuk_Vol}. 
\begin{figure}[h]{}
	\centering
	\includegraphics[width=8cm]{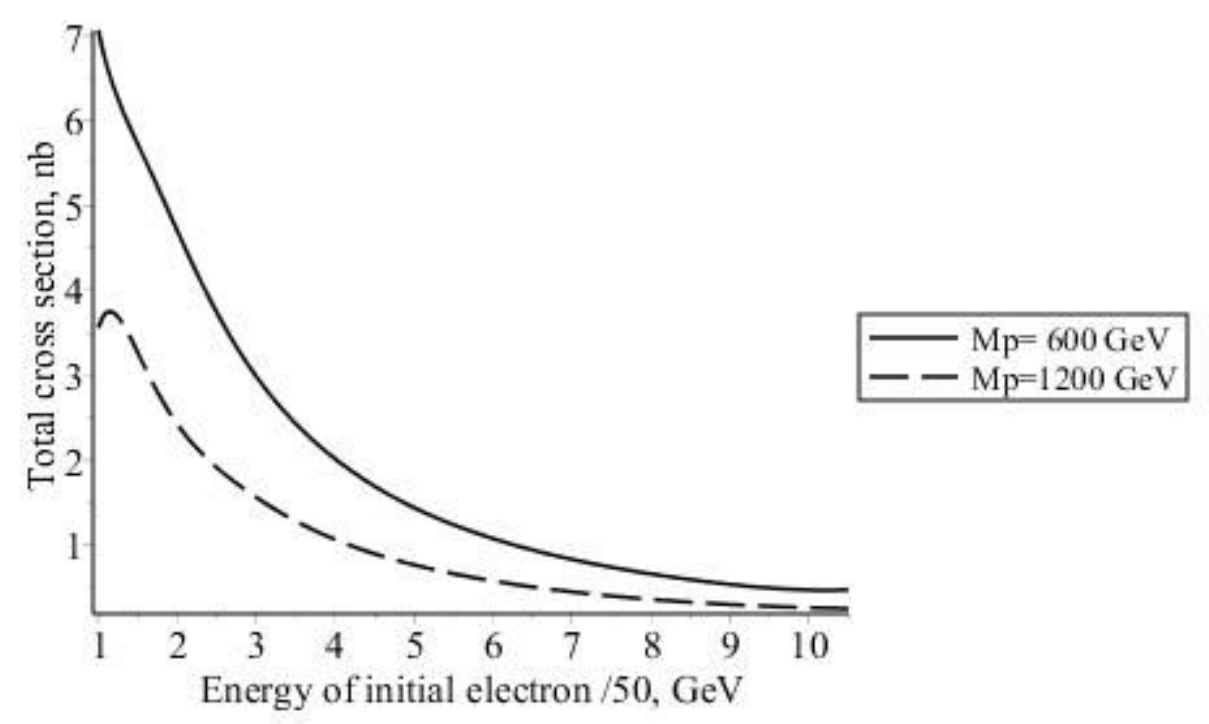}
	\caption{Total cross section of the process in dependence on the initial electron energy. For different final channels coefficients $Br_i$ should be used. Two curves correspond to $m_{\tilde \pi} =600 \, \text{and} 1200\, \mbox{GeV}$}
	\label{Tot_Cross}
\end{figure}
The cross section calculated has the form:
\begin{equation}
d\sigma(e \tilde \pi^0 \to \nu_e \tilde \pi^-) = \pi \frac{\alpha^2\cdot (1+\cos \theta)d \cos \theta}{E_e^2 \alpha_e}\cdot \frac{f_1(\alpha_e, \cos \theta)} {f_2(\alpha_e, \alpha_W, \cos \theta)},
\end{equation}
where $\alpha_e = E_e/m_{\tilde \pi^0}$, $\alpha_W = E_e/M_W$, and $E_e, \quad m_{\tilde \pi^0}, \quad M_W$ - energy of incoming electron, masses of neutral H-pion and W-boson, correspondingly. Also, we get 
$$
f_1(\alpha_e, \cos \theta) = [1-\frac{2 \cos \theta}{1+\alpha_e(1-\cos \theta)} + \frac{1}{(1+\alpha_e(1-\cos \theta))^2}]^{1/2},
$$

$$
f_2(\alpha_e, \alpha_W, \cos \theta) = [\frac{1}{\alpha_W^2} + \frac{2(1-\cos \theta)}{1+\alpha_e(1-\cos \theta)}]^2.
$$

Obviously, in the used approximation $\sigma(e \tilde \pi^0 \to \nu_e l \nu_l) = \sigma(e \tilde \pi^0 \to \nu_e \tilde \pi^-)\cdot Br(i)$ with $i=1, \, 2$.
 These coefficients should be applied for cross sections in figures below.

\begin{figure}[h]{}
	\centering
	\includegraphics[width=8cm]{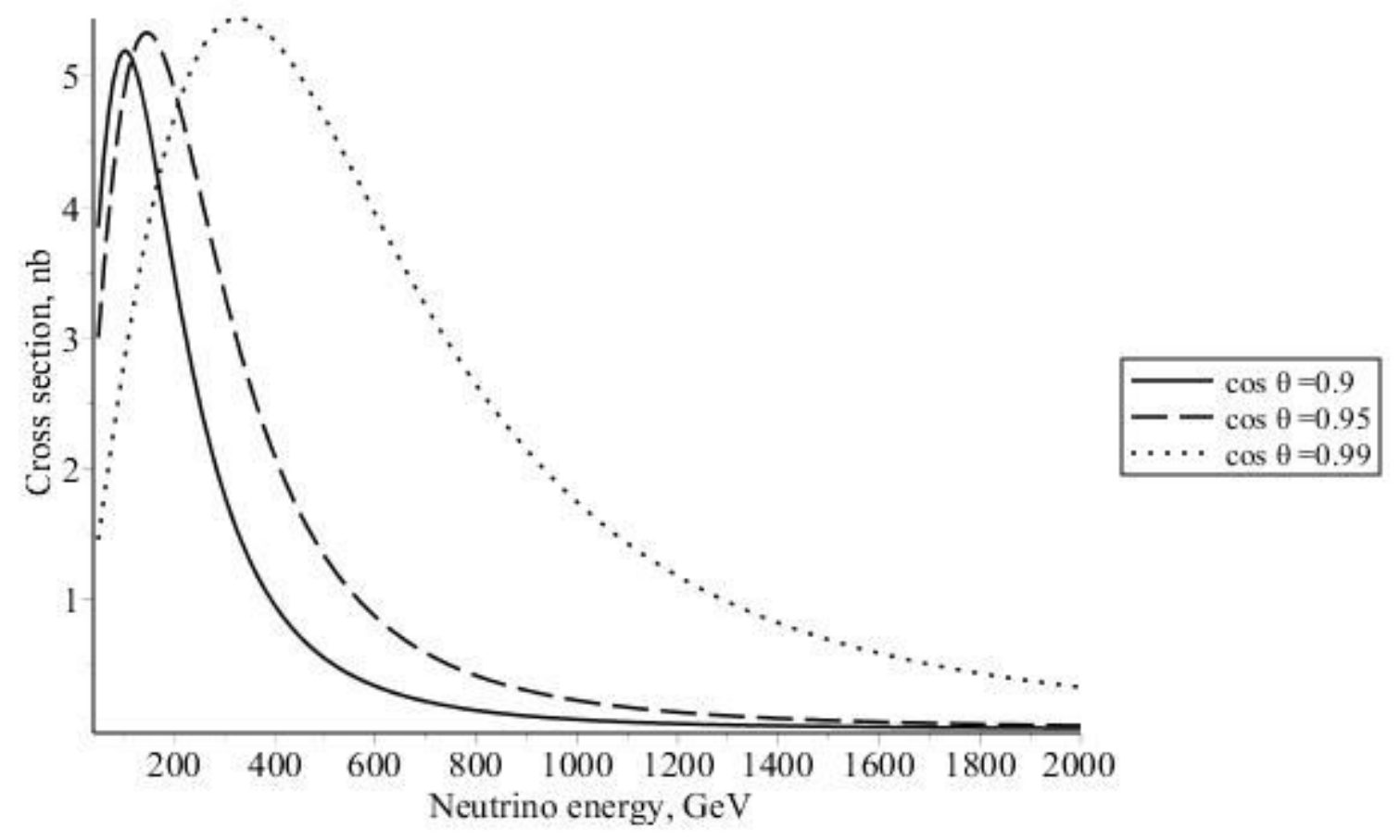}
	\caption{Depenedence of differential cross section  on the neutrino emission angle for different initial electron energies. Here  $m_{\tilde \pi} =800 \, \mbox{GeV}$}
	\label{CS-angle}
\end{figure}

\begin{figure}[h]{}
	\centering
	\includegraphics[width=8cm]{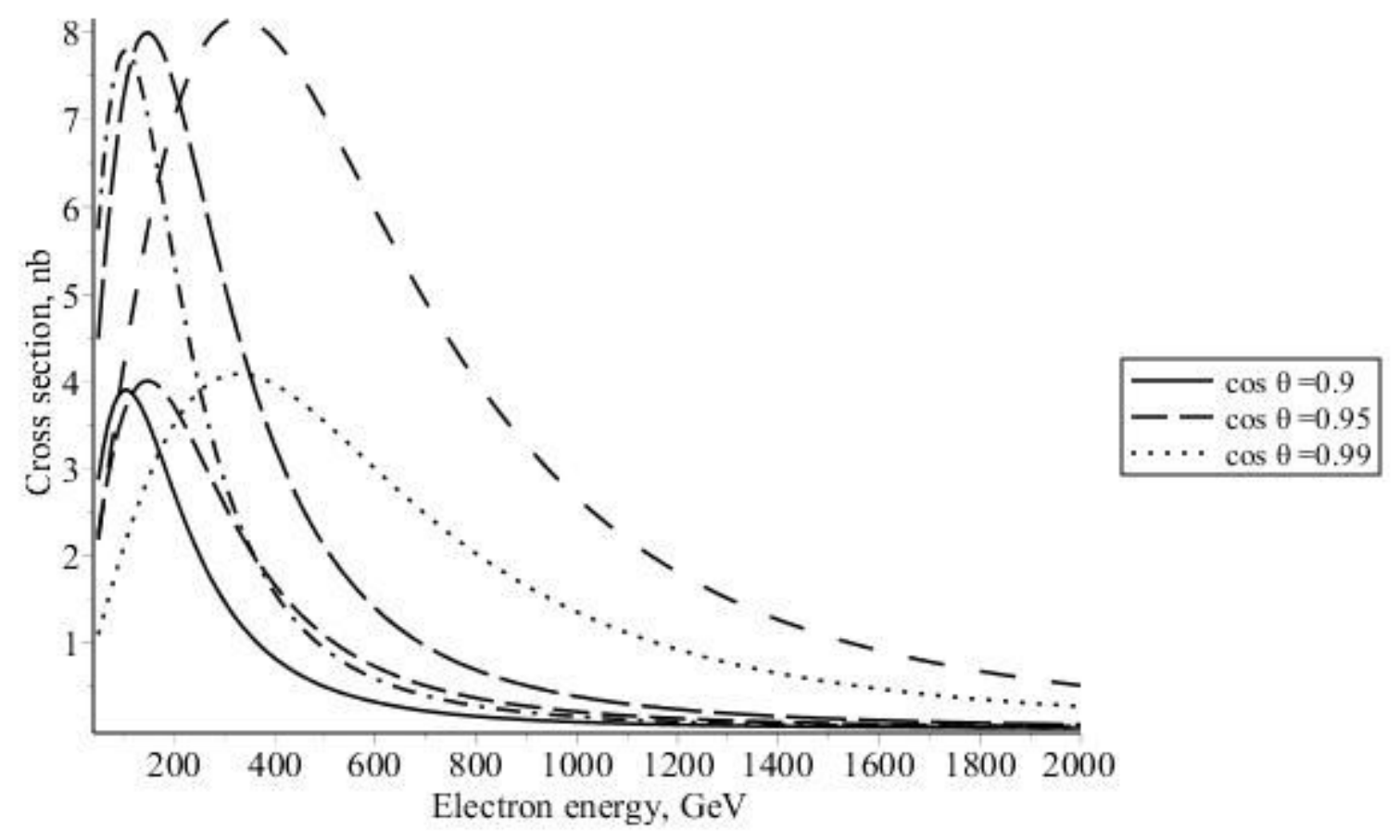}
	\caption{Dependence of differential cross section on the incident electron energy for different neutrino emission angles, Sets of curves correspond to $m_{\tilde \pi} =600 \, \mbox{GeV}$ and $1200 \, \mbox{GeV}.$}
	\label{CS_Ee}
\end{figure}
\begin{figure}[h]{}
	\centering
	\includegraphics[width=8cm]{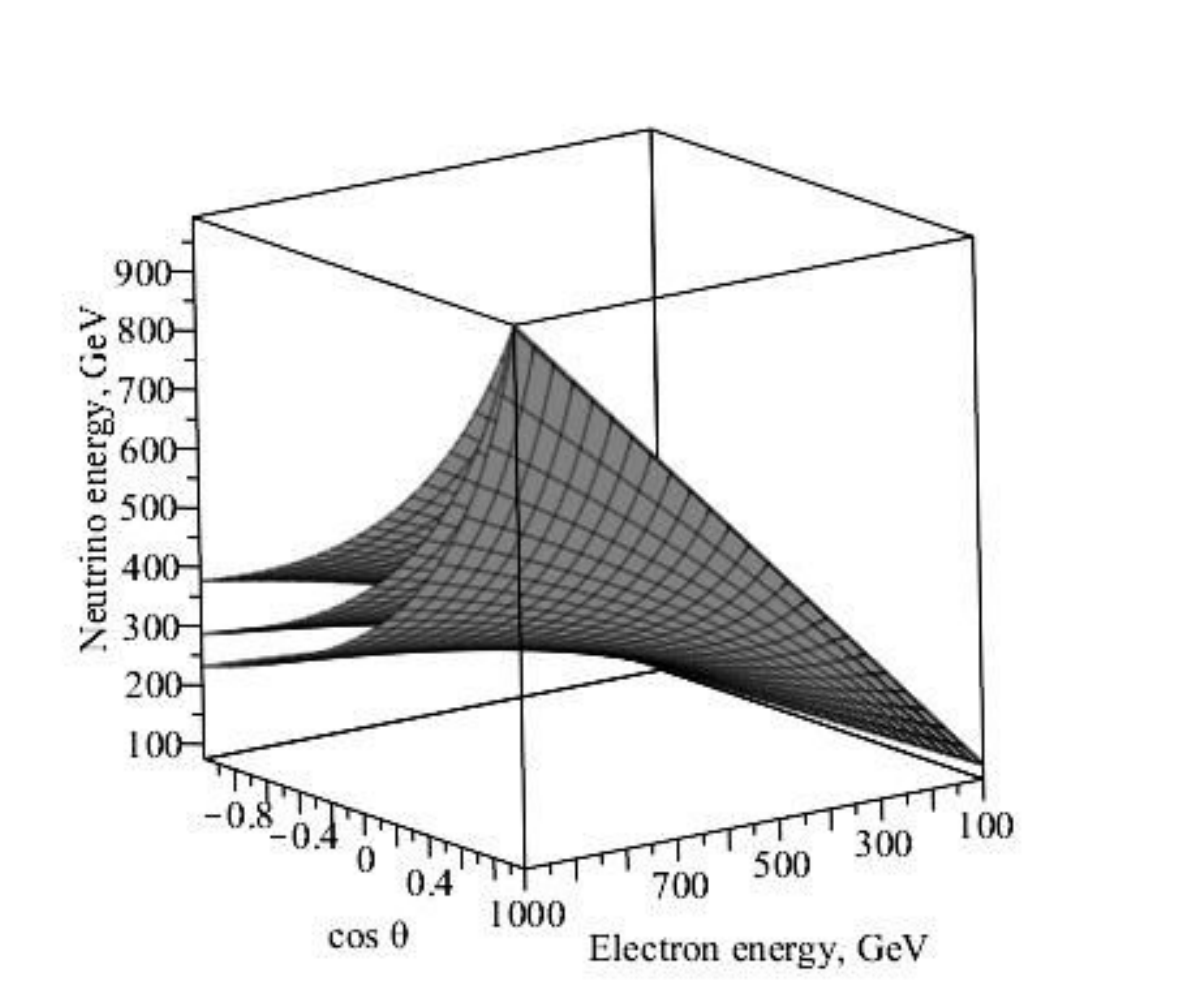}
	\caption{Dependence of neutrino energy on incident electron energy and neutrino emission angle, three lists correspond to different values of H-pion masses, $m_{\tilde \pi} =600, \,\, 800\,\, \text{and}\, 1200 \, \mbox{GeV}$}
	\label{Ee-En}
\end{figure}
Energy of secondary neutrino can be simply found from kinematics:
\begin{equation}
E_{\nu}=\frac{E_e}{1+E_e/M_{\tilde \pi}\cdot (1- \cos \theta)}.
\end{equation}
Note, there are several allowed regions for the H-pion mass resulting from the study of kinetics of the annihilation process. Here, for estimations we will use  $m_{\tilde \pi^0}=800\,\, \mbox{GeV} \,\, \text{and}\,\, 1200\,\, \mbox{GeV}$ as an average values. This allows to evaluate the effect with the sufficient accuracy. Also, the masses which are used for the process analysis are in agreement with the collider restrictions for new particles parameters.

The following figures show some of the key features of the process under analysis. Obviously, the second neutrino in the final state $\nu'_l$ has a small energy, so in these figures the parameters of the secondary neutrino, its energy and the angle at which it is emitted, refer to $\nu_e$, generated directly by the initial electron, i.e. by $e \nu W$ vertex.

From Fig.~\ref{Tot_Cross} it follows that the cross section at high energies of initial electron, $E_e = (100 - 1000) \,\, \mbox{GeV}$ decreases from $O(10) \,\,\mbox{nb}$ up to 
$O(0.1) \,\,\mbox{nb}$ and is peaked for angles of the neutrino emitting, which are close to zero. So, we have a typical picture of the forward inelastic neutrino production, the same conclusion is confirmed by Fig.~\ref{CS-angle} and Fig.~\ref{CS_Ee}.   As for energy of neutrino depending on the energy of electron, Fig.~\ref{Ee-En} demonstrates that in the approximation adopted, $E_{\nu}$ is proportional to $E_e$ and depends on the H-pion mass very slightly.

\section{Conclusions}

Some visible astrophysical phenomena can be interpreted as manifestations of the Dark Matter particles of unknown origin and nature.  Here, we consider the SM extension by minimal confined sector of hyper-quarks which have chiral symmetric interaction with standard vector bosons. In this scenario, the  $SU(4)$ symmetry breaking leads to arising of a set of pNG fields containing two stable neutral states whose mass difference is assumed as small. Thus, having the cross section of (co)annihilation of all Dark Matter components we get some  allowed regions of masses resulted from analysis of the DM components kinetics. 

The lack of any reliable collider data on New Physics including the DM nature leads to the need to look for at least some hints on the SM extension type in astrophysics. An interesting information can be extracted from studying of various astrophysical processes of production, spreading out and distributions of nuclei, particles and radiation in different regions of the Universe. Reactions with the neutrino participation provide an important data on electroweak physics at the scale of the Universe, and, moreover, participation of the DM particles in these processes  should clarify some detail of the DM nature.      

The discussed two-component model of the DM based on minimal vectorlike hypercolor, is asymmetric with respect to the electroweak interaction, namely, one of the components, $B^0$, interacts with ordinary matter only at the loop level in contrast to the neutral H-pion. Effects of inelastic scattering of cosmic rays (high-energy electrons, for example) on this H-diquark component originate from loops of heavy H-quarks and H-pions, so they are suppressed. 
Then, interaction of cosmic rays with $\tilde \pi^0$ component dominates.

We suppose that production of neutrino in the process of cosmic electron interaction with the DM components should be useful auxiliary way to analyze type and distribution of the DM in the neighborhood of the Sun and the Earth. In this case, we need in accurate measurement of angular and energy distributions of secondary neutrino fluxes.
It is important, these effects distinguish substantially from neutrino signals originating by the annihilating or decaying DM.

We can say that a weakening of energy spectrum of cosmic electrons is predicted resulting from inelastic electron scattering on hyper-pions in the hypercolor extension of the Standard Model.
In other words, high-energy electrons interacting with the H-pion DM component actually transform into electronic neutrinos, which carry away practically all the energy of primary electron. Thus, a peculiar ``burning out'' of the high-energy part of the cosmic-ray electron flux occurs. The secondary leptons (muons or electrons) occur when a charged H-pion decays, it has much less energy in the regime considered, $|t|\ll m^2_{\tilde \pi}$. Besides, two secondary neutrinos producing by different sources are significantly asymmetric in energy in this reaction, so they do not reproduce the simultaneous arising  of two neutrinos in the annihilation or decay of the DM particles.

Note, if $E_e \gg M_{\tilde \pi}$, the energy of secondary neutrino is very close to the H-pion mass, as it follows from kinematics. This effect is clearly visible at angles of neutrino emission close to $90^0$  relatively to direction of the initial electron.
Certainly, the inverse process of neutrino inelastic scatterning on the DM carriers is possible. The value of the cross section has the same order, and the kinematics is similar also.

An analysis of the lepton scattering off the DM particles in the reaction $e \tilde \pi^0 \to \nu_e \tilde \pi^0 W^-$ will be given in our forthcoming paper. Note, production of secondary high-energy neutrinos by photons interacting with the DM is of great interest and this process will be also considered. Summarizing, analysis of lepton distributions in the Universe can give an important information on the space structure and other parameters of the DM and vice versa. Moreover, specific predictions for (differential) cross sections  of such reactions depend substantially on the type of DM carriers, i.e. on the SM extension scenario. We also add that recent studies of annihilation of accumulated by the Sun DM particles with the neutrino production (see, for example, Ref.~\refcite{DMescattering}) increase the interest in consideration of the DM interaction with cosmic rays.

{\bf Acknowledgments} We thank Vladimir Korchagin and Nikolay Volchanskiy for helpful discussions and comments. M. Bezuglov is also grateful to A.N. Bezuglov for his help in computer graphics design. This work was supported by Russian Scientific Foundation (RSCF) Grant N 18-12-00213.

\end{document}